\documentclass{PoS}

\title{The hunt for axions}

\ShortTitle{The hunt for axions}

\author{Andreas Ringwald\\
        Deutsches Elektronen-Synchrotron DESY\\ Notkestr. 85\\ D-22607 Hamburg\\
        E-mail: \email{andreas.ringwald@desy.de}}

\abstract{Many theoretically well-motivated extensions of the Standard Model of particle physics 
predict the existence of the axion and further ultralight axion-like particles. They 
may constitute the mysterious dark matter in the universe and
solve some puzzles in stellar and high-energy astrophysics. There are new, relatively small 
experiments around the globe, which started to hunt for these elusive particles and
complement the accelerator based search for physics beyond the Standard Model.}

\FullConference{XVI International Workshop on Neutrino Telescopes,\\
		2-6 March 2015\\
		Palazzo Franchetti -- Istituto Veneto, Venice, Italy }

\makeatletter
\def\@oddfoot{\ifnum\thepage=1%
  \PoScopyright@box\hfill%
  \if@PoSspecialurl\PoSspecial@url\else\unhbox\PoSpaper@url\fi\fi}
\makeatother

\begin{document}

\section{Introduction}

In particle physics, we are in the comfortable situation that we have a theory -- the Standard Model (SM) 
of particle physics -- 
which describes all known particles and their interactions to a remarkable precision. On the other hand, 
astronomical observations tell us that only about fifteen percent of the matter in the universe is constituted
by the known particles. It is one of the most urgent problems of fundamental physics to understand 
the nature of the remaining eighty-five percent of matter.  

Theoretical particle physicists have been very imaginative to propose dark matter candidates 
spanning a huge range in mass and in strength of their interactions.
Among those candidates, two particular ones stick out, both because of their 
appealing physics case and because of the variety of experimental probes: the neutralino -- the lightest supersymmetric partner
of the SM neutral gauge bosons and neutral Higgses -- as a typical weakly interacting massive particle (WIMP), 
and the axion -- the pseudo Nambu-Goldstone boson arising from the breaking of a global symmetry postulated in 
order to solve the strong CP problem -- as a typical very weakly interacting slim (in the sense of ultralight)
particle (WISP). We will devote   our attention in this short review to the latter species (for more extended reviews, see Refs. \cite{Jaeckel:2010ni,Ringwald:2012hr}).

\section{Nambu-Goldstone bosons as natural WISP candidates}

WISPs occur naturally in SM extensions featuring new global $U(1)$ symmetries which are spontaneously broken by a hidden Higgs mechanism at a symmetry breaking scale $v_h$ much larger than the electroweak symmetry breaking scale, $v = 246$ GeV. In this case, the field $a(x)$ in the phase 
of the expansion of the hidden complex Higgs field about its vacuum expectation value (vev), 
$
H_h (x)= \frac{1}{\sqrt{2}}\left(v_h + h_h (x)\right) {\rm e}^{i a (x)/v_h}$,
has a flat potential, $V(a)\equiv {\rm const.}$, and is thus massless, while the field excitation 
$h_h(x)$ of the modulus  has 
a large mass $m_h\propto v_h\gg v$. Moreover, at energies below the electroweak scale, the interactions of the 
so-called Nambu-Goldstone boson $a$ with the SM particles -- gauge bosons (gluons with field strength $G$ and photons with field strength $F$) and fermions $f$ (light quarks and leptons) -- 
are suppressed by the inverse of the large symmetry breaking scale, 
\begin{equation}
{\mathcal L} = \frac{1}{2} \partial_\mu a \partial^\mu a 
- \frac{\alpha_s}{8\pi}\,C_{a g}\,\frac{a}{f_a}\,G_{\mu\nu}^c {\tilde G}^{c,\mu\nu} 
- \frac{\alpha}{8\pi} \,C_{a \gamma}\,\frac{a}{f_a}\,F_{\mu\nu} {\tilde F}^{\mu\nu} 
+ 
\frac{1}{2} \frac{C_{af}}{f_a} \, 
\partial_\mu a \ \overline{\psi}_f \gamma^\mu 
\gamma_5   \psi_f  
\label{axion_leff}
,
\end{equation}
with model-dependent dimensionless coupling coefficients  $C_{aj}$ and a decay constant $f_a\simeq v_h$.

In models in which the current associated with the new $U(1)$ symmetry features an $SU(3)_C\times SU(3)_C\times U(1)$ chiral anomaly,  $C_{ag}\neq 0$, the so-called strong CP problem is solved by non-perturba-tive QCD dynamics: the effective potential for $a$ has then an absolute minimum at $a=0$ and therefore the effective theta parameter --
proportional to the vev of $a$ -- vanishes \cite{Peccei:1977hh}. The corresponding Nambu-Goldstone boson is called the axion \cite{Weinberg:1977ma,Wilczek:1977pj}. Strictly speaking, it is a pseudo Nambu-Goldstone boson: it gets a small mass, $V(a)=\frac{1}{2} m_a^2 a^2 + {\mathcal O}(a^4)$, due to the same non-perturbative QCD effects, 
\begin{equation}
m_a = \frac{m_\pi f_\pi}{f_a/C_{ag}} \frac{\sqrt{m_u m_d}}{m_u+m_d}\simeq { 6\,  {\rm \mu eV}}
         \times
         \left(
         \frac{10^{12}\, {\rm GeV}}{f_a/C_{ag}}\right).
\label{axion_mass}
\end{equation}
Here $m_\pi$ and $f_\pi$ are the neutral pion mass and decay 
constant, and $m_u$ and $m_d$ are the masses of the light quarks. 
There may be further axion-like particles (ALPs)  which arise as Nambu-Goldstone bosons from the breaking of further well motivated global symmetries and which have no anomalous coupling to the pseudo-scalar gluon density, $C_{ag}=0$, but a non-zero anomalous coupling to photons ($C_{a\gamma}\neq 0$) and/or 
leptons ($C_{al}\neq 0$) and/or  light quarks 
($C_{aq}\neq 0$) (see, for example, \cite{Anselm:1981aw,Dias:2014osa}). In fact, string theory suggests a plenitude of such ALPs \cite{Witten:1984dg,Conlon:2006tq,Arvanitaki:2009fg,Cicoli:2012sz}.

\section{Axion/ALP dark matter?}

\begin{figure}
\begin{center}
     \includegraphics[width=.6\textwidth]{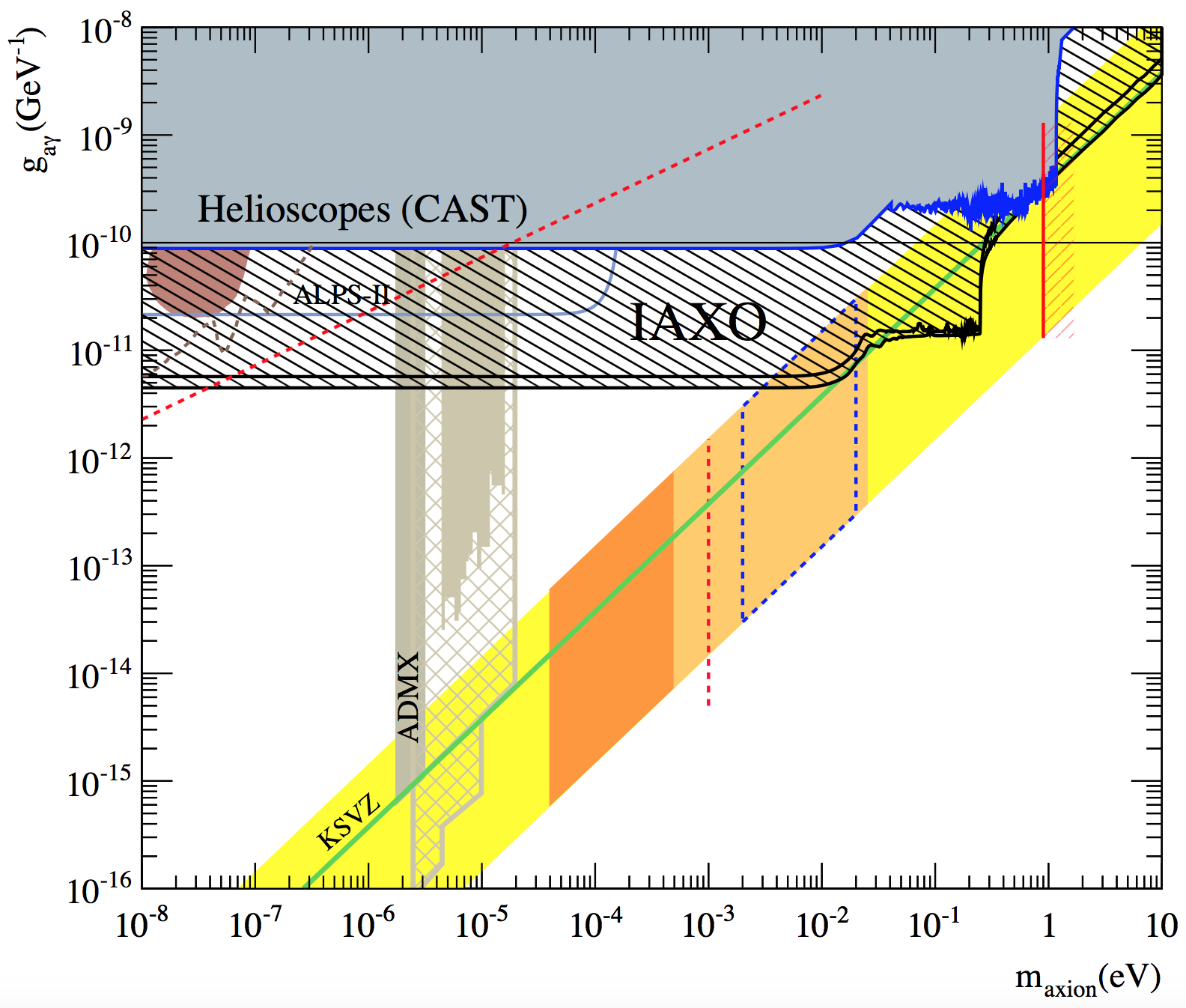}
\end{center}
     \caption{Current experimental constraints on the photon coupling of an axion or ALP \cite{Vogel:2015yka}. 
Theoretically and astrophysically favored regions are shown for axions within the yellow model band (classical axion window in dark orange, mixed axion-WIMP DM in light orange, RG and WD cooling hint within the area surrounded by the dashed blue line) and for ALPs (brown dashed line for transparency hint, below red dashed diagonal line for ALP cold DM). Future prospects of ALPS II  (above light blue line), IAXO (dashed black region), and ADMX (dashed brown region)  are also shown.
}
     \label{fig:ALP_coupling}
\end{figure}

For large symmetry breaking scales, axions and ALPs have lifetimes much longer than the age of the universe and interact extremely weakly with the SM particles, qualifying them as dark matter candidates. Moreover,  
they are produced in the early universe via the vacuum realignment mechanism as a coherent state of many, extremely non-relativistic (and thus extremely cold) particles in the form of a classical, spatially coherent oscillating field \cite{Preskill:1982cy,Abbott:1982af,Dine:1982ah}.  
Today's (time $t_0$)  fraction of axion or ALP dark matter produced via the vacuum realignment mechanism is proportional to the average field amplitude squared, $\langle a^2\rangle \equiv f_a^2 \langle \theta_a^2\rangle$,  at the time when the oscillations started, $t_{\rm osc}\simeq (3/2) \,m_a^{-1}(t_{\rm osc})$, \cite{Arias:2012az}
\begin{equation} 
\label{eq:CCDM}
R_a \equiv \frac{\rho_a}{\rho_{\rm DM}}(t_0)\simeq 0.2\,  \sqrt{\frac{m_a (t_0)}{{\rm eV}}} 
\sqrt{\frac{m_a (t_0)}{m_a(t_{\rm osc})}}
\left(\frac{f_a}{10^{11}\, {\rm GeV}}\right)^2  \langle \theta_a^2\rangle\ 
 .
\end{equation}
Here, the indicated time-dependence of the mass arises from its temperature dependence, $m_a (t)\equiv m_a (T(t))$, taking into account possible 
plasma effects. From (\ref{eq:CCDM}) it follows, that an appreciable fraction of light axion/ALP dark matter is only expected for sufficiently large symmetry breaking scale, $f_a\gtrsim 10^{10 - 12}$ GeV and correspondingly small couplings to SM particles (see Fig. \ref{fig:ALP_coupling}).  

\section{Hints on axion/ALPs from astrophysics}

\subsection{Axion/ALP energy losses of stars?}

Axions or ALPs could be produced in hot stellar plasmas and could thus transport energy out of stars, thereby possibly influencing stellar lifetimes and energy-loss rates. 
Stringent bounds on the axion or ALP couplings to photons and electrons in a wide mass range 
have been established from observations of stars in globular-clusters (GCs) \cite{Raffelt:1996wa}. 
These are relying on the fact that number counts of stars in particular branches of the color-magnitude diagram of
GCs allow for 
detailed tests of stellar evolution. Recently, several authors have confronted new data with improved theoretical predictions and found hints for anomalous excessive energy losses.  

A recent state-of-the-art analysis of Horizontal Branch (i.e. helium burning) stars (HBs) in a large sample of 39 Galactic GCs, exploiting modern stellar models, 
and taking into account the substantial dependence of the predictions on the He mass fraction,  
found a slight indication of additional losses which may be accounted by 
Primakoff-like axion/ALP emission in the Coulomb field of a charged particle, 
$
\gamma+Ze\rightarrow Ze+a$,
if the photon coupling is in the range \cite{Ayala:2014pea}
\begin{eqnarray}
\label{Eq:1sigma}
g_{a \gamma} \equiv \frac{\alpha}{2\pi} \frac{C_{a\gamma}}{f_a}= 4.5_{-1.6}^{+1.2} \times 10^{-11}~{\rm GeV}^{-1} 
\,\ ,
\hspace{6ex} {\rm for\ }\ m_a\lesssim {\rm few}\times 10\ {\rm keV}. 
\end{eqnarray}
Conservatively, the authors of this analysis determine also an 
upper bound, $g_{a\gamma}<6.6\times 10^{-11}\,\textrm{GeV}^{-1}$, at 95 \% confidence level (CL),
which represents the strongest limit on $g_{a\gamma}$ for a wide mass range. 

Clearly, the result  (\ref{Eq:1sigma}) gives only  a marginal hint for the existence of the QCD axion or an ALP. In fact, at around two sigma it is  still compatible with the SM, $g_{a\gamma}=0$. 
However, there are other mismatches between theory and observations which may also be seen as slight indications for the existence of a Nambu-Goldstone boson with such a coupling to the photon. 
One of them is the fact that the ratio of blue to red supergiants (SGs) is smaller than predicted by 
stellar evolution models. Moreover, the blue SGs appear to be less blue than expected \cite{McQuinn:2011bb}. 
This effect could also be explained by an axion/ALP with 
$g_{a\gamma}={\rm few}\times 10^{-11}\,{\rm GeV}^{-1}$ 
 \cite{Friedland:2012hj,Carosi:2013rla}. 

Interestingly enough,  
Red Giants (RGs) in GCs also mildly prefer additional energy losses, this time, however, due to axion/ALP emission via bremsstrahlung 
 of axions or ALPs,
$
e+Ze\to Ze+e+a$,
pointing to an electron coupling in the range  
 \cite{Viaux:2013lha}
\begin{eqnarray}
\label{Eq:1sigma_RG}
g_{a e}\equiv \frac{C_{ae} m_e}{f_a} = 1.8_{-0.8}^{+0.6} \times 10^{-13} 
\,\ ,
\hspace{6ex} {\rm for\ }\ m_a\lesssim {\rm few}\times 10\ {\rm keV}.
\end{eqnarray}
Still, at $\sim 2\sigma$, the result is compatible with the SM, 
$g_{ae} < 4.3\times 10^{-13}\, (95\% \,\ \textrm{CL})$.

Another astrophysical observable probing the electron coupling is the luminosity function 
of white dwarfs (WDs). Intriguingly, recent analyses, based on detailed WD cooling treatment and new 
data, find weak evidence that the WD luminosity 
function fits better with a new energy-loss channel that can be 
interpreted in terms of axion/ALP losses via bremsstrahlung in electron-ion
or electron-electron collsions, requiring an electron coupling in the range 
$
g_{ae}= 1.0^{+0.2}_{-0.2}\times 10^{-13}$, for 
$m_a\lesssim {\rm few}\times 10\ {\rm keV}$  \cite{Isern:2008nt,Bertolami:2014wua}.
This is consistent with the axion/ALP explanation of the apparent
excessive energy loss of RGs, cf. Eq. (\ref{Eq:1sigma_RG}). However, here again  
the evidence is still rather weak: at $2\sigma$ the WD luminosity function is consistent
with the SM,
$g_{ae} < 2.3\times 10^{-13}\,(95\% \,\ \textrm{CL})$. 

Finally, there is another hint for extra energy losses from the neutron star in Cassiopeia A (Cas A):
its surface temperature measured  over 10 years reveals an unusually fast cooling rate. This 
may be interpreted as a hint 
for extra cooling by axion/ALP due to nucleon bremsstrahlung, 
$
N+N \rightarrow N+N+a$,
requiring a 
coupling to the neutron of size  \cite{Leinson:2014ioa}
\begin{equation}
\label{eq:hint_nscasa}
g_{an}\equiv \frac{C_{an} m_n}{f_a} \sim  4\times 10^{-10}, \hspace{6ex} {\rm for\ }\ 
m_a\lesssim {\rm few}\times 10\ {\rm MeV}.
\end{equation}
Still, this can also be seen as an approximate upper limit on this coupling. In fact, as recently pointed out, the more rapid cooling of the superfluid core in the neutron star may also arise from a phase transition of the neutron condensate into a multicomponent state \cite{Leinson:2014cja}.

Clearly, the evidence for each of the anomalous energy loss hints is very weak. But nevertheless, 
a unified picture seems to emerge: just one light, $m_a\lesssim {\rm few}\times 10$ keV, pseudo Nambu-Goldstone boson
with 
$
f_a\sim 10^8\ {\rm GeV}$, $C_{a\gamma}\sim 1$, and $C_{ae}\sim C_{an}\sim 10^{-2}$,
is required to 
explain at the same time the indications from the three quite different stellar energy loss channels. 

Intriguingly, in LARGE volume string compactifications, in which the volume $\mathcal V$ of the compactified extra dimensions is stabilized at an exponentially large value in units of the string size,
the decay constant of closed string ALPs is generically much smaller than the Planck scale,  $f_a \sim M_{\rm Pl}/\sqrt{\mathcal V}$, and their matter coupling coefficients $C_{af}$   are generically 
suppressed by a factor $\alpha\sim 10^{-2}$ in comparison to the photon coupling coefficient $C_{a\gamma}$, 
realizing the required properties  \cite{Cicoli:2012sz}.

\subsection{Photon-ALP oscillations in astrophysical magnetic fields?}

Photon-ALP oscillations in large-scale interstellar as well as in local magnetic fields associated with astrophysical 
objects provide another sensitive probe for ALPs complementary to Primakoff conversion  in stars. 
However, this probe is only effective for large photon-ALP mixing and is therefore confined to very low masses, 
\begin{equation}
\mid m_a^2 - \omega_{\rm pl}^2\mid \ll 2 g_{a\gamma}\, B\, E\simeq 2.5\times {\rm neV}^2 
\left( \frac{g_{a\gamma}}{\rm 10^{-11}\ GeV^{-1}}\right) 
\left( \frac{B}{\rm \mu G}\right)
\left( \frac{E}{\rm GeV}\right)
,
\end{equation}
where $\omega_{\rm pl}\simeq 4\times 10^{−11} {\rm eV} \sqrt{􏰒n_e/{\rm cm^{−3}}}$ is the plasma frequency
in terms of the electron density in the medium, while $B$ is the magnetic field strength and $E$ is the 
photon/axion/ALP energy.

In a galactic core-collapse supernova (SN), ALPs would be emitted via the Primakoff process, and oscillate into gamma rays in the magnetic field of the Milky Way. 
The lack of a gamma-ray signal in the GRS instrument of the SMM satellite in coincidence with the observation of the neutrinos emitted from SN1987A therefore provides a strong bound on their coupling to photons, cf.  \cite{Brockway:1996yr,Grifols:1996id}. Recently, this bound has been revisited and the underlying physics has been brought 
to the current state-of-the-art, as far as modelling of the supernova 
and the Milky-Way magnetic field are concerned, resulting in the limit   \cite{Payez:2014xsa}
\begin{equation}
g_{a\gamma}< 􏰂5.3\times 10^{-12}\  {\rm GeV}^{-1},\hspace{6ex}  {\rm for\ }\  
m_a\lesssim  􏰂4.4\times 10^{-10}\ {\rm eV}. 
\end{equation}
Therefore, the mass window for ALPs explaining the HB energy loss shrinks to 
 $\rm neV\lesssim m_a\lesssim few\times 10\  keV$.

Gamma-ray spectra from distant active galactic nuclei (AGN) should show
an energy and redshift-dependent exponential attenuation, $\exp (-\tau (E,z))$,
due to $e^+ e^-$ pair production off the extragalactic background light (EBL) -- the stellar and
dust-reprocessed light accumulated during the cosmological evolution following
the era of re-ionization. 
However, a number of authors \cite{De Angelis:2007dy,Horns:2012fx,Rubtsov:2014uga} have noted that  the observed spectra seem to point to an anomalous transparency of the universe for gamma-rays at large optical depth, $\tau\gtrsim 2$, although with evidence below two sigma, if one takes into account systematic effects 
(EBL spectrum, intrinsic source spectra, etc.) \cite{Biteau:2015xpa}.
This may be explained by photon $\leftrightarrow$ ALP oscillations:
the conversion of gamma rays into ALPs in the
magnetic fields around AGNs or in the intergalactic medium, followed by their unimpeded
travel towards our galaxy and the consequent reconversion into photons in the (inter)galactic magnetic
fields,
require an ALP with \cite{Mirizzi:2007hr,Simet:2007sa,SanchezConde:2009wu,Mirizzi:2009aj,Horns:2012kw,Meyer:2013pny,Tavecchio:2014yoa}
\begin{equation}
 g_{a\gamma}\gtrsim  10^{-(11 - 12)} \ {\rm GeV}^{-1},\hspace{6ex} {\rm  for\ }\  
 m_{a}\lesssim 10^{-7}\  {\rm eV}.
\end{equation}

Intriguingly, a similar value of the photon coupling 
is required also for
a possible explanation of the rapidly varying very high energy ($E>50$ GeV) emission from the flat spectrum radio quasar PKS 1222+216  which represents a challenge for standard blazar scenarios: in the latter one
is forced to invoke the existence of a very compact emitting region at a large distance from the jet base, in order to avoid absorption of gamma rays in the dense ultraviolet radiation field of the broad line region. 
In ref.~\cite{Tavecchio:2012um} it was shown that one can also use a
standard blazar model for PKS 1222+216 where gamma rays are produced close to the central engine,
if one assumes that inside the source photons can oscillate into ALPs.
Moreover, the required photon coupling overlaps with the preferred region from the anomalous energy 
losses of helium burning starts in globular clusters.

Finally, it was found that observed soft X-ray excesses in many galaxy clusters may be explained by the conversion of 
a cosmic ALP background (CAB) radiation, corresponding to an effective number 
$\triangle N_{\rm eff}$ of extra neutrinos, into photons in the cluster magnetic fields \cite{Conlon:2013txa,Angus:2013sua,Kraljic:2014yta}.
This explanation requires
that the CAB spectrum is peaked in the soft X-ray region and that the ALP coupling and mass satisfy
$
g_{a\gamma}\gtrsim  (1 - 2)\times 10^{-13} \ {\rm GeV}^{-1}\, \sqrt{0.5/\triangle N_{\rm eff}}$, for 
$m_a\lesssim 10^{-12}\ {\rm eV}$.

\section{The experimental hunt for the axion and ALPs}

We have seen that there is a strong physics case for the axion and other ALPs arising as Nambu-Goldstone 
bosons from the breaking of a symmetry at a scale $f_a\sim 10^{8\div 12}$ GeV. Fortunately, such WISPs 
are in reach of a number of experiments which are presently carried out or being set up.

\subsection{Light-shining-through-a-wall searches}

Light-Shining-Through-Wall (LSW) experiments aim both for production and detection of axions and ALPs in the laboratory. 
This is done by sending laser photons along a strong magnetic field, allowing for their conversion into axions or ALPs, towards a blocking wall, behind of which the latter may then reconvert, 
again in a strong magnetic field, into photons, the latter being susceptible to detection (for a review, see \cite{Redondo:2010dp}). 
The Any Light Particle Search (ALPS) 
experiment at DESY and OSQAR at CERN share currently the best sensitivity of LSW experiments 
\cite{Ehret:2010mh,Pugnat:2013dha}. 
ALPS II \cite{Bahre:2013ywa} proposes to use 10+10 straightened 
HERA magnets \cite{Ringwald:2003nsa}, a high-power laser
system developed and the pioneering realization of an optical
regeneration cavity \cite{Hoogeveen:1990vq,Sikivie:2007qm} developed by the Albert-Einstein Institute, and a superconducting low-background detector. 
It aims at crucially testing the ALP explanation of the excessive helium burning star energy loss and of the 
anomalous cosmic gamma ray transparency,  cf. Fig.~\ref{fig:ALP_coupling}. 

\subsection{Helioscope searches}

Helioscopes aim at detecting solar axions and ALPs produced by their conversion into photons inside of a strong magnet pointing towards the Sun \cite{Sikivie:1983ip}. The CERN Axion Solar Telescope (CAST), employing an LHC dipole test magnet,  currently sets the best helioscope 
limit \cite{Andriamonje:2007ew,Arik:2013nya}.
A proposed next-generation axion helioscope, dubbed the International Axion Observatory (IAXO),
envisions a dedicated superconducting toroidal magnet with much bigger aperture than CAST,
a detection system consisting of large X-ray telescopes coupled to
ultra-low background X-ray detectors, and a large, robust tracking system \cite{Armengaud:2014gea}. It aims at the 
sensitivity shown in Fig.~\ref{fig:ALP_coupling}. 
It will crucially test the axion explanation of the energy losses of helium burning stars, red giants, white dwarfs and the neutron star in Cas A, and of the ALP explanation of the anomalous cosmic gamma ray transparency. Futhermore, it will also scratch the parameter region required to explain the soft X-ray excess from galaxy clusters. 

\subsection{Direct axion/ALP dark matter searches}

\subsubsection{Haloscopes}

Haloscopes directly search for  galactic halo dark matter axions and ALPs in the laboratory via their coupling to the photon. 
Currently, the most sensitive ones exploit electromagnetic cavites placed in  
a strong magnet \cite{Sikivie:1983ip}. They aim for the detection of electromagnetic power arising from
the conversion of dark matter axions or ALPs into real photons, with frequency 
$
\nu=m_{a}/(2\pi )=0.24\ {\rm GHz}\times (m_{a}/\mu{\rm eV})$.
The best sensitivity is reached on resonance, the power output then being proportional to the quality factor of the cavity. 
The Axion Dark Matter eXperiment (ADMX) has indeed reached recently the sensitivity to probe 
axion dark matter \cite{Asztalos:2009yp} (see Fig. \ref{fig:ALP_coupling}).
Further microwave cavity based haloscope opportunities in complementary mass ranges 
may arise from recycling available microwave cavities and magnets at accelerator laboratories \cite{Baker:2011na,Horns:2014qta}. 
 
Other new concepts for haloscopes are also being investigated \cite{Sikivie:2013laa}.  
A microwave Fabry-Perot resonator in a spatially varying magnetic field may be exploited  to search for axion/ALP dark matter
with masses above 40 $\mu$eV \cite{Rybka:2014cya}. 
Converted photons from axion/ALP dark matter could be focused in a manner similar to a dish antenna, allowing for broad-band 
searches \cite{Horns:2012jf}. 

\subsubsection{Magnetic resonance searches}

Axion dark matter produced via the misalignment mechanism gives oscillating electric dipole moments (EDMs)
to all nucleons. These  EDMs cause the precession of nuclear spins in a nucleon spin polarized
sample in the presence of an electric field. The 
resulting transverse magnetization can be searched for by exploiting magnetic-resonance (MR) techniques
 \cite{Budker:2013hfa}. The aim of the corresponding Cosmic Axion Spin Precession Experiment (CASPEr) in Mainz
is to probe axion dark matter in the anthropic window, corresponding to GUT to Planck scale 
symmetry-breaking scales $f_a\sim 10^{15 - 18}$ GeV, complementary to the classic axion window probed by  ADMX. 

The axion/ALP nucleon coupling $C_{aN}$ will also lead to a spin precession about the axion/ALP DM wind, even without 
the presence of an electric field. Therefore, CASPEr can also be exploited to search for the 
magnetization due to this effect  \cite{Graham:2013gfa}. 
Unfortunately, the projected sensitivity does not reach the axion prediction in this case.

Finally, the axion/ALP electron coupling $C_{ae}$ will also lead to a spin precession about the axion/ALP DM wind 
\cite{Krauss:1985ub,Barbieri:1985cp}. 
The QUAX (QUaerere AXions) experiment in preparation in Italy by INFN aims at exploiting MR  
inside a magnetized material \cite{QUAX}. 
Because of the higher Larmor frequency of the electron, it extends the sensitivity to higher masses.

\section{Summary}

There is a strong physics case for the axion and ALPs.  They occur naturally 
in many 
theoretically appealing ultraviolet completions of the SM. 
They are dark matter
candidates and can explain the hints for an excessive energy loss of stars and for an anomalous transparency of the universe for TeV photons. 
A significant portion of their parameter space will be tackled in this 
decade by experiments. Stay tuned!


\begin{thebibliography}{99}

\bibitem{Jaeckel:2010ni}
  J.~Jaeckel and A.~Ringwald,
  ``The Low-Energy Frontier of Particle Physics,''
  Ann.\ Rev.\ Nucl.\ Part.\ Sci.\  {\bf 60} (2010) 405
  [arXiv:1002.0329 [hep-ph]].

\bibitem{Ringwald:2012hr}
  A.~Ringwald,
  ``Exploring the Role of Axions and Other WISPs in the Dark Universe,''
  Phys.\ Dark Univ.\  {\bf 1} (2012) 116
  [arXiv:1210.5081 [hep-ph]].

\bibitem{Peccei:1977hh}
  R.~D.~Peccei and H.~R.~Quinn,
  ``CP Conservation in the Presence of Instantons,''
  Phys.\ Rev.\ Lett.\  {\bf 38} (1977) 1440.

\bibitem{Weinberg:1977ma}
  S.~Weinberg,
  ``A New Light Boson?,''
  Phys.\ Rev.\ Lett.\  {\bf 40} (1978) 223.

\bibitem{Wilczek:1977pj}
  F.~Wilczek,
  ``Problem of Strong P and T Invariance in the Presence of Instantons,''
  Phys.\ Rev.\ Lett.\  {\bf 40} (1978) 279.

\bibitem{Anselm:1981aw}
  A.~A.~Anselm and N.~G.~Uraltsev,
  ``A Second Massless Axion?,''
  Phys.\ Lett.\ B {\bf 114} (1982) 39.


\bibitem{Dias:2014osa}
  A.~G.~Dias,  A.~C.~B.~Machado, C.~C.~Nishi, A.~Ringwald and P.~Vaudrevange,
  ``The Quest for an Intermediate-Scale Accidental Axion and Further ALPs,''
  JHEP {\bf 1406} (2014) 037 
  [arXiv:1403.5760 [hep-ph]].

\bibitem{Witten:1984dg}
  E.~Witten,
  ``Some Properties of O(32) Superstrings,''
  Phys.\ Lett.\ B {\bf 149} (1984) 351.

\bibitem{Conlon:2006tq}
  J.~P.~Conlon,
  ``The QCD axion and moduli stabilisation,''
  JHEP {\bf 0605} (2006) 078
  [hep-th/0602233].

\bibitem{Arvanitaki:2009fg}
  A.~Arvanitaki, S.~Dimopoulos, S.~Dubovsky, N.~Kaloper and J.~March-Russell,
  ``String Axiverse,''
  Phys.\ Rev.\ D {\bf 81} (2010) 123530
  [arXiv:0905.4720 [hep-th]].

\bibitem{Cicoli:2012sz}
  M.~Cicoli, M.~Goodsell and A.~Ringwald,
  ``The type IIB string axiverse and its low-energy phenomenology,''
  JHEP {\bf 1210} (2012) 146
  [arXiv:1206.0819 [hep-th]].

\bibitem{Preskill:1982cy}
  J.~Preskill, M.~B.~Wise and F.~Wilczek,
  ``Cosmology of the Invisible Axion,''
  Phys.\ Lett.\ B {\bf 120} (1983) 127.

\bibitem{Abbott:1982af}
  L.~F.~Abbott and P.~Sikivie,
  ``A Cosmological Bound on the Invisible Axion,''
  Phys.\ Lett.\ B {\bf 120} (1983) 133.

\bibitem{Dine:1982ah}
  M.~Dine and W.~Fischler,
  ``The Not So Harmless Axion,''
  Phys.\ Lett.\ B {\bf 120} (1983) 137.

\bibitem{Arias:2012az}
  P.~Arias, D.~Cadamuro, M.~Goodsell, J.~Jaeckel, J.~Redondo and A.~Ringwald,
  ``WISPy Cold Dark Matter,''
  JCAP {\bf 1206} (2012) 013
  [arXiv:1201.5902 [hep-ph]].

\bibitem{Vogel:2015yka}
  J.~K.~Vogel {\it et al.}  
[IAXO Collaboration],
  ``The Next Generation of Axion Helioscopes: The International Axion Observatory (IAXO),''
  Phys.\ Procedia {\bf 61} (2015) 193.

\bibitem{Raffelt:1996wa}
  G.~G.~Raffelt,
  ``Stars as laboratories for fundamental physics: The astrophysics of neutrinos, axions, and other weakly interacting particles,''
  Chicago, USA: Univ. Pr. (1996) 664 p.

\bibitem{Ayala:2014pea}
  A.~Ayala, I.~Dominguez, M.~Giannotti, A.~Mirizzi and O.~Straniero,
  ``Revisiting the bound on axion-photon coupling from Globular Clusters,''
  Phys.\ Rev.\ Lett.\  {\bf 113} (2014) 191302 
  [arXiv:1406.6053 [astro-ph.SR]].

\bibitem{McQuinn:2011bb}
  K.~B.~W.~McQuinn, E.~D.~Skillman, J.~J.~Dalcanton, A.~E.~Dolphin, J.~Holtzman, D.~R.~Weisz and B.~F.~Williams,
  ``Observational Constraints on Red and Blue Helium Burning Sequences,''
  Astrophys.\ J.\  {\bf 740} (2011) 48
  [arXiv:1108.1405 [astro-ph.CO]].

\bibitem{Friedland:2012hj}
  A.~Friedland, M.~Giannotti and M.~Wise,
  ``Constraining the Axion-Photon Coupling with Massive Stars,''
  Phys.\ Rev.\ Lett.\  {\bf 110} (2013) 6,  061101
  [arXiv:1210.1271 [hep-ph]].

\bibitem{Carosi:2013rla}
  G.~Carosi,  A.~Friedland, M.~Giannotti, M.~J.~Pivovaroff, J.~Ruz and J.~K.~Vogel,
  ``Probing the axion-photon coupling: phenomenological and experimental perspectives. A snowmass white paper,''
  arXiv:1309.7035 [hep-ph].

\bibitem{Viaux:2013lha}
  N.~Viaux, M.~Catelan, P.~B.~Stetson, G.~Raffelt, J.~Redondo, A.~A.~R.~Valcarce and A.~Weiss,
  ``Neutrino and axion bounds from the globular cluster M5 (NGC 5904),''
  Phys.\ Rev.\ Lett.\  {\bf 111} (2013) 231301
  [arXiv:1311.1669 [astro-ph.SR]].

\bibitem{Isern:2008nt}
  J.~Isern, E.~Garcia-Berro, S.~Torres and S.~Catalan,
  ``Axions and the cooling of white dwarf stars,''
  Astrophys.\ J.\  {\bf 682} (2008) L109
  [arXiv:0806.2807 [astro-ph]].

\bibitem{Bertolami:2014wua}
  M.~M.~Miller Bertolami, B.~E.~Melendez, L.~G.~Althaus and J.~Isern,
  ``Revisiting the axion bounds from the Galactic white dwarf luminosity function,''
  JCAP {\bf 1410} (2014) 069
  [arXiv:1406.7712 [hep-ph]].

\bibitem{Leinson:2014ioa}
  L.~B.~Leinson,
  ``Axion mass limit from observations of the neutron star in Cassiopeia A,''
  JCAP {\bf 1408} (2014) 031
  [arXiv:1405.6873 [hep-ph]].

\bibitem{Leinson:2014cja}
  L.~B.~Leinson,
  ``Superfluid phases of triplet pairing and rapid cooling of the neutron star in Cassiopeia A,''
  Phys.\ Lett.\ B {\bf 741} (2015) 87
  [arXiv:1411.6833 [astro-ph.SR]].

\bibitem{Brockway:1996yr}
  J.~W.~Brockway, E.~D.~Carlson and G.~G.~Raffelt,
  ``SN1987A gamma-ray limits on the conversion of pseudoscalars,''
  Phys.\ Lett.\ B {\bf 383} (1996) 439
  [astro-ph/9605197].

\bibitem{Grifols:1996id}
  J.~A.~Grifols, E.~Masso and R.~Toldra,
  ``Gamma-rays from SN1987A due to pseudoscalar conversion,''
  Phys.\ Rev.\ Lett.\  {\bf 77} (1996) 2372
  [astro-ph/9606028].

\bibitem{Payez:2014xsa}
  A.~Payez, C.~Evoli, T.~Fischer, M.~Giannotti, A.~Mirizzi and A.~Ringwald,
  ``Revisiting the SN1987A gamma-ray limit on ultralight axion-like particles,''
  JCAP {\bf 1502} (2015) 006
  [arXiv:1410.3747 [astro-ph.HE]].

\bibitem{De Angelis:2007dy}
  A.~De Angelis, M.~Roncadelli and O.~Mansutti,
  ``Evidence for a new light spin-zero boson from cosmological gamma-ray propagation?,''
  Phys.\ Rev.\ D {\bf 76} (2007) 121301
  [arXiv:0707.4312 [astro-ph]].

\bibitem{Horns:2012fx}
  D.~Horns and M.~Meyer,
  ``Indications for a Pair-Production Anomaly from the Propagation of VHE Gamma-Rays,"
  JCAP {\bf 1202} (2012) 033
  [arXiv:1201.4711 [astro-ph.CO]].

\bibitem{Rubtsov:2014uga}
  G.~I.~Rubtsov and S.~V.~Troitsky,
  ``Breaks in gamma-ray spectra of distant blazars and transparency of the Universe,''
  JETP Lett.\  {\bf 100} (2014) 397
   [JETP Lett.\  {\bf 100} (2014) 355]
  [arXiv:1406.0239 [astro-ph.HE]].

\bibitem{Biteau:2015xpa}
  J.~Biteau and D.~A.~Williams,
  ``The extragalactic background light, the Hubble constant, and anomalies: conclusions from 20 years of TeV gamma-ray observations,''
  arXiv:1502.04166. 

\bibitem{Mirizzi:2007hr} 
 A.~Mirizzi, G.~G.~Raffelt and P.~D.~Serpico,
 ``Signatures of axion-like particles in the spectra of TeV gamma-ray sources,''
 Phys.\ Rev.\ D {\bf 76} (2007) 023001
 [arXiv:0704.3044 [astro-ph]].

\bibitem{Simet:2007sa}
  M.~Simet, D.~Hooper and P.~D.~Serpico,
  ``The Milky Way as a Kiloparsec-Scale Axionscope,''
  Phys.\ Rev.\ D {\bf 77} (2008) 063001
  [arXiv:0712.2825 [astro-ph]].

\bibitem{SanchezConde:2009wu}
  M.~A.~Sanchez-Conde {\it et al.},  
  ``Hints of the existence of Axion-Like-Particles from the gamma-ray spectra of cosmological sources,''
  Phys.\ Rev.\ D {\bf 79} (2009) 123511
  [arXiv:0905.3270 [astro-ph.CO]].

\bibitem{Mirizzi:2009aj} 
 A.~Mirizzi and D.~Montanino,
 ``Stochastic conversions of TeV photons into axion-like particles in extragalactic magnetic fields,''
 JCAP {\bf 0912} (2009) 004
 [arXiv:0911.0015 [astro-ph.HE]].

\bibitem{Horns:2012kw} 
 D.~Horns, L.~Maccione, M.~Meyer, A.~Mirizzi, D.~Montanino and M.~Roncadelli,
 ``Hardening of TeV gamma spectrum of AGNs in galaxy clusters by conversions of photons into axion-like particles,''
 Phys.\ Rev.\ D {\bf 86} (2012) 075024
 [arXiv:1207.0776 [astro-ph.HE]].

\bibitem{Meyer:2013pny}
  M.~Meyer, D.~Horns and M.~Raue,
  ``First lower limits on the photon-axion-like particle coupling from very high energy gamma-ray observation,''
  Phys.\ Rev.\ D {\bf 87} (2013) 035027
  [arXiv:1302.1208 [astro-ph.HE]].

\bibitem{Tavecchio:2014yoa}
  F.~Tavecchio, M.~Roncadelli and G.~Galanti,
  ``Photons to axion-like particles conversion in Active Galactic Nuclei,''
  Phys.\ Lett.\ B {\bf 744} (2015) 375
  [arXiv:1406.2303 [astro-ph.HE]].

\bibitem{Tavecchio:2012um}
  F.~Tavecchio, M.~Roncadelli, G.~Galanti and G.~Bonnoli,
  ``Evidence for an axion-like particle from PKS 1222+216?,''
  Phys.\ Rev.\ D {\bf 86} (2012) 085036
  [arXiv:1202.6529 [astro-ph.HE]].

\bibitem{Conlon:2013txa}
  J.~P.~Conlon and M.~C.~D.~Marsh,
  ``Excess Astrophysical Photons from a 0.1-1 keV Cosmic Axion Background,''
  Phys.\ Rev.\ Lett.\  {\bf 111} (2013) 151301
  [arXiv:1305.3603 [astro-ph.CO]].

\bibitem{Angus:2013sua}
  S.~Angus, J.~P.~Conlon, M.~C.~D.~Marsh, A.~J.~Powell and L.~T.~Witkowski,
  ``Soft X-ray Excess in the Coma Cluster from a Cosmic Axion Background,''
  JCAP {\bf 1409} (2014) 026 
  [arXiv:1312.3947 [astro-ph.HE]].

\bibitem{Kraljic:2014yta}
  D.~Kraljic, M.~Rummel and J.~P.~Conlon,
  ``ALP Conversion and the Soft X-ray Excess in the Outskirts of the Coma Cluster,''
  JCAP {\bf 1501} (2015) 011
  [arXiv:1406.5188 [hep-ph]].
 
\bibitem{Redondo:2010dp}
  J.~Redondo and A.~Ringwald,
  ``Light shining through walls,''
  Contemp.\ Phys.\  {\bf 52} (2011) 211
 [arXiv:1011.3741 [hep-ph]].
 
\bibitem{Ehret:2010mh}
  K.~Ehret {\it et al.} [ALPS Collaboration],  
  ``New ALPS Results on Hidden-Sector Lightweights,''
  Phys.\ Lett.\ B {\bf 689} (2010) 149 
   [arXiv:1004.1313 [hep-ex]].

\bibitem{Pugnat:2013dha}
  P.~Pugnat {\it et al.}  [OSQAR Collaboration],
  ``Search for weakly interacting sub-eV particles with the OSQAR laser-based experiment: results and perspectives,''
  Eur.\ Phys.\ J.\ C {\bf 74} (2014) 8,  3027
  [arXiv:1306.0443 [hep-ex]].

\bibitem{Bahre:2013ywa}
  R.~B\"ahre
{\it et al.} [ALPS Collaboration],
  ``Any light particle search II - Technical Design Report,''
  JINST {\bf 8} (2013) T09001
  [arXiv:1302.5647 [physics.ins-det]].

\bibitem{Ringwald:2003nsa}
  A.~Ringwald,
  ``Production and detection of very light bosons in the HERA tunnel,''
  Phys.\ Lett.\ B {\bf 569} (2003) 51
  [hep-ph/0306106].

\bibitem{Hoogeveen:1990vq}
  F.~Hoogeveen and T.~Ziegenhagen,
  ``Production and detection of light bosons using optical resonators,''
  Nucl.\ Phys.\ B {\bf 358} (1991) 3.

\bibitem{Sikivie:2007qm}
  P.~Sikivie, D.~B.~Tanner and K.~van Bibber,
  ``Resonantly enhanced axion-photon regeneration,''
  Phys.\ Rev.\ Lett.\  {\bf 98} (2007) 172002 
  [hep-ph/0701198 [HEP-PH]].

\bibitem{Sikivie:1983ip}
  P.~Sikivie,
  ``Experimental Tests of the Invisible Axion,''
  Phys.\ Rev.\ Lett.\  {\bf 51} (1983) 1415
   [Erratum-ibid.\  {\bf 52} (1984) 695].

\bibitem{Andriamonje:2007ew}
  S.~Andriamonje {\it et al.}  [CAST Collaboration],
  ``An Improved limit on the axion-photon coupling from the CAST experiment,''
  JCAP {\bf 0704} (2007) 010
  [hep-ex/0702006].

\bibitem{Arik:2013nya}
  M.~Arik {\it et al.}  [CAST Collaboration],  
  ``CAST solar axion search with $^3$He buffer gas: Closing the hot dark matter gap,''
  Phys.\ Rev.\ Lett.\  {\bf 112} (2014) 091302 
  [arXiv:1307.1985 [hep-ex]].

\bibitem{Armengaud:2014gea}
  E.~Armengaud {\it et al.}   [IAXO Collaboration],  
  ``Conceptual Design of the International Axion Observatory (IAXO),''
  JINST {\bf 9} (2014) T05002
   [arXiv:1401.3233 [physics.ins-det]].

\bibitem{Asztalos:2009yp}
  S.~J.~Asztalos {\it et al.}  [ADMX Collaboration],
  ``A SQUID-based microwave cavity search for dark-matter axions,''
  Phys.\ Rev.\ Lett.\  {\bf 104} (2010) 041301
   [arXiv:0910.5914 [astro-ph.CO]].

\bibitem{Baker:2011na}
  O.~K.~Baker, M.~Betz, F.~Caspers, J.~Jaeckel, A.~Lindner, A.~Ringwald, Y.~Semertzidis and P.~Sikivie {\it et al.},
  ``Prospects for Searching Axion-like Particle Dark Matter with Dipole, Toroidal and Wiggler Magnets,''
  Phys.\ Rev.\ D {\bf 85} (2012) 035018
  [arXiv:1110.2180 [physics.ins-det]].

\bibitem{Horns:2014qta}
  D.~Horns, A.~Lindner, A.~Lobanov and A.~Ringwald,
  ``WISP Dark Matter eXperiment and Prospects for Broadband Dark Matter Searches in the $1\,\mu$eV--$10\,$meV Mass Range,''
  arXiv:1410.6302 [hep-ex].

\bibitem{Sikivie:2013laa}
  P.~Sikivie, N.~Sullivan and D.~B.~Tanner,
  ``Proposal for Axion Dark Matter Detection Using an LC Circuit,''
  Phys.\ Rev.\ Lett.\  {\bf 112} (2014) 13,  131301
  [arXiv:1310.8545 [hep-ph]].

\bibitem{Rybka:2014cya}
  G.~Rybka,  A.~Wagner, A.~Brill, K.~Ramos, R.~Percival and K.~Patel,
  ``Search for dark matter axions with the Orpheus experiment,''
  Phys.\ Rev.\ D {\bf 91} (2015) 1,  011701
  [arXiv:1403.3121 [physics.ins-det]].

\bibitem{Horns:2012jf}
  D.~Horns, J.~Jaeckel, A.~Lindner, A.~Lobanov, J.~Redondo and A.~Ringwald,
  ``Searching for WISPy Cold Dark Matter with a Dish Antenna,''
  JCAP {\bf 1304} (2013) 016
  [arXiv:1212.2970].

\bibitem{Budker:2013hfa}
  D.~Budker,  P.~W.~Graham, M.~Ledbetter, S.~Rajendran and A.~Sushkov,
  ``Cosmic Axion Spin Precession Experiment (CASPEr),''
  Phys.\ Rev.\ X {\bf 4} (2014) 021030
  [arXiv:1306.6089 [hep-ph]].

\bibitem{Graham:2013gfa}
  P.~W.~Graham and S.~Rajendran,
  ``New Observables for Direct Detection of Axion Dark Matter,''
  Phys.\ Rev.\ D {\bf 88} (2013) 035023
  [arXiv:1306.6088 [hep-ph]].

\bibitem{Krauss:1985ub}
  L.~Krauss, J.~Moody, F.~Wilczek and D.~E.~Morris,
  ``Calculations for Cosmic Axion Detection,''
  Phys.\ Rev.\ Lett.\  {\bf 55} (1985) 1797.

\bibitem{Barbieri:1985cp}
  R.~Barbieri, M.~Cerdonio, G.~Fiorentini and S.~Vitale,
  ``Axion To Magnon Conversion: A Scheme for The Detection of Galactic Axions,''
  Phys.\ Lett.\ B {\bf 226} (1989) 357.

\bibitem{QUAX}
  G.~Carugno, A.~Ortolan, G.~Ruoso and C.~Speake,
  ``QUaerere AXion -  A proposal for a search of galactic axions using magnetized materials,''
  October 8, 2014. 


\end{thebibliography}
\end{document}